\documentclass[aps, prl,onecolumn,10pt,showpacs,superscriptaddress,amsmath,amssymb,nobibnotes]{revtex4-1}

\usepackage{physics}
\usepackage{graphicx}% Include figure files
\usepackage{dcolumn}% Align table columns on decimal point
\usepackage{bm}% bold math
\usepackage{amssymb}      
\usepackage{subfigure}  
 \usepackage{booktabs}
\usepackage{natbib}
\usepackage{color}
\begin{document}

\title{Security Analysis and Improvement of Source Independent Quantum \\Random Number Generators with Imperfect Devices}% Force line breaks with \\

	\author{Xing Lin}
	\author{Shuang Wang}
	\email{wshuang@ustc.edu.cn}
\author{Zhen-Qiang Yin}
\author{Guan-Jie Fan-Yuan}
\author{Rong Wang}
\author{Wei Chen}
\author{De-Yong He}
\author{Zheng Zhou}
\author{Guang-Can Guo}
\author{Zheng-Fu Han}
\affiliation{CAS Key Laboratory of Quantum Information, University of Science and Technology of China, Hefei, Anhui 230026, China}
\affiliation{CAS Center for Excellence in Quantum Information and Quantum Physics, University of Science and Technology of China, Hefei, Anhui 230026, China}
\affiliation{State Key Laboratory of Cryptology, P. O. Box 5159, Beijing 100878, P. R. China}%Lines break automatically or can be forced with \\
%\affiliation{%
%\affiliation{
%\affiliation{
%\affiliation{%
%             %  but any date may be explicitly specified

\begin{abstract}
A quantum random number generator (QRNG) as a genuine source of randomness is essential in many applications, such as number simulation and cryptography. Recently, a source-independent quantum random number generator (SI-QRNG), which can generate secure random numbers with untrusted sources, has been realized. However, the measurement loopholes of the trusted but imperfect devices used in SI-QRNGs have not yet been fully explored, which will cause security problems, especially in high-speed systems. Here, we point out and evaluate the security loopholes of practical imperfect measurement devices in SI-QRNGs. We also provide corresponding countermeasures to prevent these information leakages by recalculating the conditional minimum entropy and adding a monitor. Furthermore, by taking into account the finite-size effect,we show that the influence of the afterpulse can exceed that of the finite-size effect with the large number of sampled rounds. Our protocol is simple and effective, and it promotes the security of SI-QRNG in practice as well as the compatibility with high-speed measurement devices, thus paving the way for constructing ultrafast and security-certified commercial SI-QRNG systems.

%\begin{description}
%\end{description}
\end{abstract}

\maketitle

%\tableofcontents

\noindent{\bf INTRODUCTION}

\noindent Random numbers have become a core element in many fields, ranging from daily applications, such as lotteries, to scientific simulation and cryptography. Pseudo or classical random number generators, relying on deterministic algorithms or physical processes, have been widely used. However, their predictability and strong long-range correlation mean that they are not suitable for applications that need high security, such as cryptography.

In contrast, quantum random number generators (QRNGs) are considered the best solution for generating unpredictable random numbers by exploiting the intrinsic uncertainty of quantum mechanics\cite{born1926quantenmechanik}. Many QRNG protocols have been presented recently; they are based on different sources, such as the spatial\cite{stefanov2000optical,jennewein2000fast,grafe2014chip,oberreiter2016light,wang2006scheme} and temporal\cite{dynes2008high,wayne2009photon,wahl2011ultrafast,nie2014practical} modes of photons, vacuum-state fluctuations\cite{gabriel2010generator,shen2010practical,symul2011real,zhu2012unbiased,zhou2019practical}, laser phase noise\cite{qi2010high,guo2010truly,jofre2011true,zhou2015randomness}, stimulated scattering\cite{bustard2011quantum,england2014efficient} and other quantum phenomena\cite{zhou2019quantum,demir2020security,yao2018multi,vallone2014quantum,ma2016quantum}. Most of them rely on fully trusted devices, however, realistic devices are usually imperfect or even untrusted, and they might provide side information to eavesdroppers and cause overestimation of the conditional min-entropy.

Device-independent QRNGs (DI-QRNGs), which are based on the violation of the Bell inequality, have been proposed to solve this problem, but their extremely low bit rates and low loss tolerance limit their development\cite{pironio2010random,christensen2013detection,bierhorst2018experimentally}. To date, the rate of the fastest DI-QRNG has been reported to be 181 bps\cite{liu2018device}, which is very low for practical applications. To increase both the bit rate and security, source independent QRNGs (SI-QRNGs), as a compromise solution, have been proposed\cite{cao2016source,xu2016experimental,marangon2017source,avesani2018source,PhysRevX.10.041048,zheng2020bias}. By a proper conditional min-entropy estimation, SI-QRNGs can generate high-speed secure random numbers with the untrusted source.

In general, measurement devices in SI-QRNGs are trusted and cannot be controlled by attackers. However, practical measurement devices are not perfect. It is reasonable to suppose that there is no classical or quantum relevance between measurement devices and attackers, but we do not prevent attackers from obtaining the essential parameters of detectors that can reveal the detector imperfections. These imperfections, such as the afterpulse, provide side information to attackers and thus impact the security of SI-QRNGs.

In this work, we build a model of practical imperfect measurement devices and evaluate the influences of these imperfections on the discrete-variable SI-QRNG. We also propose a protocol to eliminate these influences and then estimate the conditional min-entropy and the rate of this SI-QRNG. In the framework of the discrete variable SI-QRNGs, single photon avalanche detectors (SPDs) are the core detection components and the imperfect factors of practical SPDs, such as the afterpulse, detector efficiency and sensitivity to photon number distribution, will cause the conditional min-entropy to be estimated incorrectly. Here, we present an effective method to estimate the conditional min-entropy focusing on the afterpulse. Additionally, we analyse the influences of detector efficiency mismatch and photon number distribution in entropy estimation and then consider a scheme to remove them. Finally, by using the random sampling method as the previous protocol\cite{cao2016source} and entropy inequality method\cite{tomamichel2012tight}, we analyze the secure randomness rates with the finite key effect and compare the influences of different factors.

This paper is organized as follows. First, we describe how random numbers can be generated by a typical SI-QRNG. Then, we show the effects of different parameters and present a model to account for these effects. In addition, we analyse a numerical simulation of the finite size effect. Finally, we conclude with a discussion.

\hfill

\noindent{\bf RESULTS}

\noindent{\textit{A typical SI-QRNG}}

\noindent In a discrete-variable SI-QRNG scenario, as shown in Fig.1, the source is an untrusted party that might be controlled by an attacker, Eve, and Alice has trusted measurement devices such as threshold detectors (SPDs), a polarizing beam splitter (PBS) and a filter as well as the trusted randomness encoding device. By estimating the conditional min-entropy based on the error rate of the X-basis measurement, randomness extraction can then extract uniform random numbers from the original data\cite{cao2016source}. The detailed process is as follows.
\begin{figure}
	\centering
	\includegraphics[width=1\linewidth]{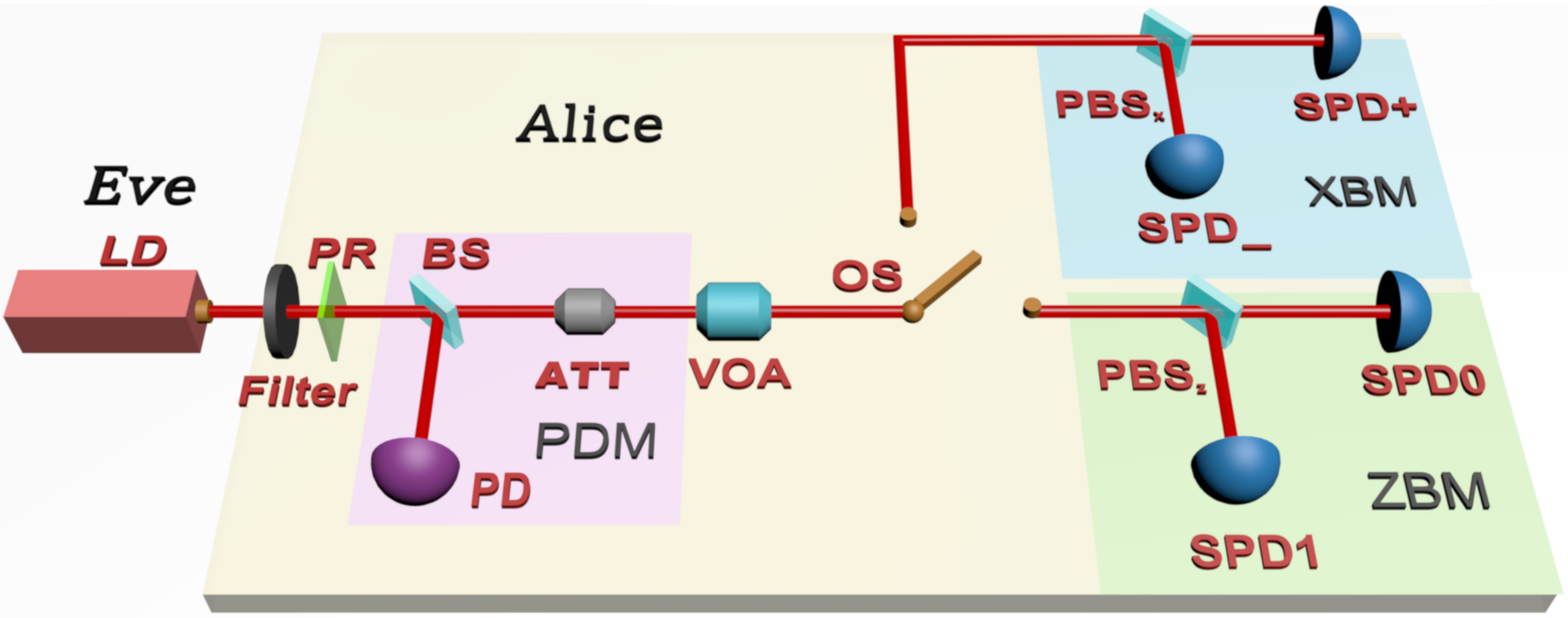}
	\caption{A schematic diagram of a SI-QRNG. Eve controls the laser diode (LD) and sends pulses with states that are changed to mixed states through a filter and a phase randomizer (PR). Through a photon distribution monitor (PDM) which consists of a beam splitter (BS), an attenuator (ATT), a photodiode (PD) and a variable optical attenuator (VOA), the pulses are sent to an optical switch (OS) to choose the measurement basis-the X basis measurement (XBM) or Z basis measurement (ZBM)-both of which consist of a polarization beam splitter (PBS) and two single photon detectors (SPD). The PDM block is used for the distribution monitor and does not exist in general SI-QRNGs in \cite{cao2016source}}
	\label{fig:fig1}
\end{figure}

First, the untrusted source which might be controlled by Eve emits $ N $ pulses with quantum state $ \rho $. From the perspective of Alice, the emitted photons should be in the qubit state $ \ket{+}=(\ket{0}+\ket{1})/\sqrt{2}$.

After proper filtering and attenuation, Alice randomly chooses $ n_{x} $ pulses and measures them in the X basis to estimate error. The remaining $ n_{z}=N-n_{x} $ pulses will be used to generate raw random numbers in the Z basis measurement.

Note that it is a key assumption that the measurement devices are compatible with the squashing model\cite{beaudry2008squashing}. In an ideal scheme, a pulse with multiphotons will be \textit{squashed} into a qubit; therefore, the unknown arbitrary-dimensional signal state emitted from the source will become the qubit or vacuum state. Then, $ n_{x}^{'} $ qubits in the X basis and $ n_{z}^{'} $ qubits in the Z basis will be detected with post-selection of the vacua. In practice, the threshold detectors are considered equivalents of the squashing operation in this protocol; therefore we can use the squashing model directly in the analysis. However, the threshold detectors are usually placed at the end of the system, and a double click might not be avoided. Therefore, in the postprocess, it is necessary to randomly assign the outcome to 0 or 1 for double-click events to satisfy the needs of the squashing model\cite{gittsovich2014squashing}.

According to the error rate $ e_{bx} $, which represents the ratio of detecting $ \ket{-}=(\ket{0}-\ket{1})/\sqrt{2} $ in the X basis, and the complementary uncertainty relation, as well as that in the quantum key distribution (QKD)\cite{shor2000simple}, we can obtain the extractable random numbers from the raw data: $ K=n_{z}^{'}[1-h(e_{bx}+\theta)]-t_{e} $, where $ \theta $ is the deviation due to statistical fluctuations, $ 2^{-t_{e}} $ is the failure probability of the randomness extraction and $ h(x) $ represents the binary Shannon entropy function of $ x $. 

Finally, uniform random numbers can be obtained from the raw data by a randomness extractor. A random seed with a length of $ n_{seed} $ and $ n_{post} $ will be consumed in the basis choice and the postprocess, respectively.

\hfill

\noindent{\textit{Model}}

\noindent In the previous SI-QRNG scheme, the untrusted source was the focus of the QRNG, and previous work has tried to eliminate Eve's influence on the source. However, in real implementation, the ignored imperfections of measurement devices, such as the afterpulse, detection efficiency mismatch and sensitivity to the photon number distribution, also have a strong impact on security. In what follows, we first build the underlying response probability model with these factors and recalculate the entropy to reveal these factors' influence mechanism on SI-QRNGs. Then we propose a scheme to solve these problems.

For threshold detectors, SPDs, the usual model of the response probabilities without the afterpulse effect $ p_{\alpha} ^{d}$ can be written as\cite{yu2016reexamination}
\begin{eqnarray}
~~~~~~~~~~~p^{d}_{\alpha}=1-\tau_{\alpha}(1-e_{d\alpha}) ~~~~~~~~\alpha\in\{0,1,+,-\},~\label{1}
\end{eqnarray}
where $\tau_{\alpha} $ is the zero photon distribution probability after the influence of the loss and detector efficiency and $ e_{d\alpha} $ is the background counting rate. The subscript $ \alpha $ indicates the different detectors which include $ D_{0} $ and $ D_{1} $ in the Z basis and $ D_{+} $ and $ D_{-} $ in the X basis. 

One of the most important factors of SPDs is the afterpulse, which has considerable effects on high-speed systems and will be blinding existing analytical models from reality. Therefore, it is necessary to build an afterpulse-compatible model. To adapt the model with the afterpulse, we should change the response probabilities $ p_{\alpha} ^{d}$ to
\begin{eqnarray}
p_{\alpha}=1-\tau_{\alpha}(1-e_{d\alpha})(1-P_{ap\alpha}),\label{2}
\end{eqnarray}
where $ P_{ap\alpha} $ is the practical current afterpulse probability of each detector which is given by\cite{fan2018afterpulse}
\begin{eqnarray}
	 P_{ap\alpha}=\frac{\hat{p_{\alpha}}}{1-\hat{p_{\alpha}}}p^{d}_{b\alpha}, \label{3}
\end{eqnarray}
where $ \hat{p_{\alpha}}= \sum^{n}_{j=1}\hat{p}_{j\alpha} $ is the overall first-order afterpulse rate, and $ \hat{p}_{j\alpha} $ is the first-order afterpulse coefficient contributed by the former \textit{j}th detection window avalanche. $ p^{d}_{b\alpha} $ represents the former response ratio without the afterpulse. It should be noted that we only consider infinite former responses here and the finite responses influence will be discussed later.

Furthermore, to achieve compatibility between our model and the squashing model, it is necessary to precisely depict the probabilities of single-click and double-click events. Moreover, we define the error counts as the events in which only $ D_{-} $ responds, and according to the squashing model, a double click event should be assigned a random bit, which will add a half error count in the X basis. Therefore, according to Eq.\ref{2}, the probabilities of single-click and double-click events $ Q_{single} $, $ Q_{double} $ in the Z basis and the error rate $ EQ $ in the X basis are given by
	\begin{eqnarray}
	&&Q_{single}=p_{0}(1-p_{1})+p_{1}(1-p_{0}), \label{4a}
	\\
	&&Q_{double}=p_{0}p_{1}, \label{4b}
	\\
	&&EQ=p_{-}(1-p_{+})+\frac{1}{2}p_{-}p_{+}. \label{4c}
	\end{eqnarray}

\hfill

\noindent{\textit{Afterpulse}}

The afterpulse is a key factor in SPD, especially in high-speed systems. In a discrete variable QRNG system, as the afterpulse probability becomes very high with the increase in the system speed, it has a significant impact on not only the random number generation rate but also on the security, which has been ignored in previous works. 

Here, we first consider the afterpulse influence mechanism on raw random numbers generated on the Z basis. Intuitively, for a dual detector system, we can infer that the afterpulse will increase the response probability of the detector and increase the production probability of '00' or '11'. This means that the raw sequence will come to have more positive correlations, which will lead to the overestimation of the entropy and thus an information leak to Eve in the previous models. In what follows, we quantitatively analyse the influence of afterpulse on random number sequences.

In statistical analysis, we always use the autocorrelation coefficient $ a_{i} $ to describe the \textit{i}th autocorrelation of a n-bit sequence $ \{x_{i}\} $\cite{knuth1997art}:
\begin{eqnarray}
a_{i}=\frac{\sum_{j=1}^{n-i}(x_{j}-\bar{x})(x_{j+i}-\bar{x})}{\sum_{j=1}^{n}(x_{j}-\bar{x})^{2}},\label{5}
\end{eqnarray}
where $ \bar{x} $ is the expectation value of $ \{x_{i}\} $. In general, if the sequence is a series of random numbers with good statistical characteristics, the theoretical expectation of $ a_{i} $ should be 0.

With the model presented in the previous sections, we can derive the \textit{prior} autocorrelation coefficient with afterpulse $ a^{p}_{i} $:
\begin{eqnarray}
a^{p}_{i}=&&[p^{i1}_{1}(1-p^{i0}_{0})-p^{i0}_{1}(1-p^{i1}_{0})](1-k)\nonumber\\
	&&+[p^{i0}_{0}(1-p^{i1}_{1})-p^{i1}_{0}(1-p^{i0}_{1})](-k),
\end{eqnarray}
where $ p^{i1}_{\alpha} $ is the response probability of $ D_{\alpha} $ with the former \textit{i}th detection response and  $ p^{i0}_{\alpha} $ is that without the former \textit{i}th detection response. $ k $ is the expectation of this raw sequence, which is given by
\begin{eqnarray}
k=\frac{p_{1}(1-p_{0})}{p_{1}(1-p_{0})+p_{0}(1-p_{1})}.
\end{eqnarray}

Note that we assign double-click events to random bits, so these events should not affect the autocorrelation coefficient in the statistical analysis. Therefore, here we only consider the random numbers generated by single-click events.

\begin{figure}
	\centering
	\includegraphics[width=0.9\linewidth]{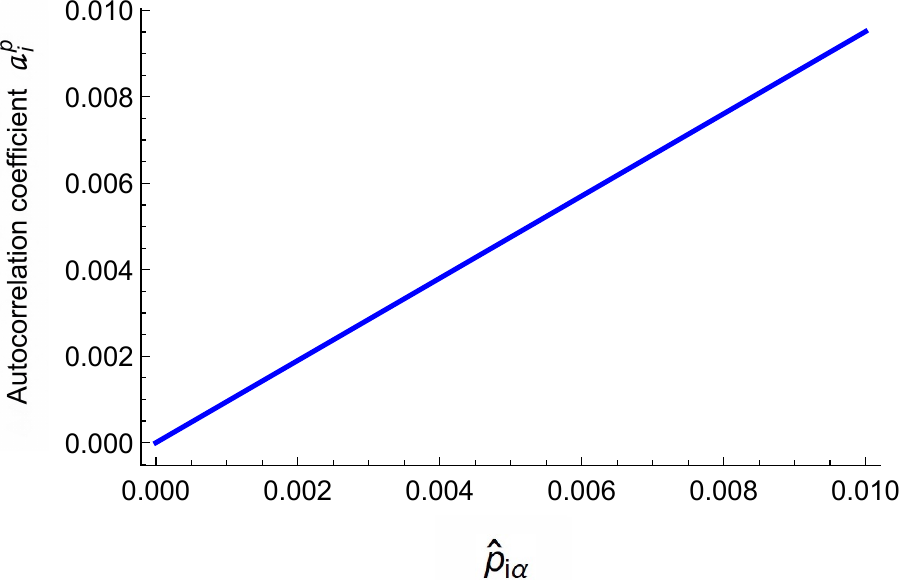}
	\caption{The relation function between $ a^{p}_{i} $ and $\hat{p}_{i\alpha}$. The same values are given to the parameters of the detectors. We assume that pulses with coherent states that contain the mean photon number $ \nu=1 $ insert SPDs with detection efficiency $ \eta_{\alpha}=0.1 $, $ e_{d\alpha}=6\times10^{-7} $ and $ \hat{p_{\alpha}}=0.05 $. For ideal devices, the \textit{prior} autocorrelation coefficient should be 0. }
	\label{fig:api}
\end{figure}

In the METHODS section, we will show the detail of the $ a^{p}_{i} $ calculation and it is proven that the relation between $ a^{p}_{i} $ and $\hat{p}_{i\alpha}$ is quadratic and degenerates to linear when all the parameters of the two detectors are the same. As shown in Fig.~\ref{fig:api}, with the increase of $\hat{p}_{i\alpha}$, $ a^{p}_{i} $ increases rapidly and has an obviously positive correlation. 

To solve this security issue, it is necessary to analyse the entropy adapted to the afterpulse model. Here, we define $ H_{min}(A|E) $ as the total conditional minimum entropy of the raw random numbers. It needs to be emphasized that the use of minimum entropy is necessary because Eve is allowed to obtain the probability distribution $ \{P_{i}\} $ of the raw sequence, and there is an optimum strategy for Eve that she can use to guess the maximum probability event and obtain more information than under the limitation of conditional entropy. Therefore, we should estimate this worst-case and calculate the minimum randomness event with the afterpulse.

We first consider only the model of raw randomness generation in the Z basis and calculate the conditional minimum entropy $ H_{min}(Z|E) $ with the afterpulse under this condition. According to the response probability model with the afterpulse in Eqs.~(\ref{2}) and (\ref{3}), the afterpulse probability $ P_{ap\alpha} $ is dependent on the former response ratio $ p^{d}_{b\alpha} $. Let us consider the maximum leakage information condition, in which the contribution of $ P_{ap\alpha} $ to the two detectors reaches its maximum difference. If $ D_{0} $ always responds before, which is opposite to the behaviour of $ D_{1} $, the afterpulse contributions to the two detectors will become the most unbalanced. Understandably, the final distribution $ P(i) $ will deviate far from Alice's estimation and Eve will obtain the maximum amount of leaked information. Under this condition, the response probabilities $ p_{0} $ and $ p_{1} $ will become:
\begin{eqnarray}
&&p_{0}^{(1)}=1-\tau_{0}(1-e_{d0})(1-\frac{\hat{p_{0}}}{1-\hat{p_{0}}}),\nonumber\\
&&p_{1}^{(0)}=1-\tau_{1}(1-e_{d1}).
\end{eqnarray}
Assuming that the input state is $ \ket{+} $, we can obtain the conditional minimum entropy $ H_{min}(Z|E) $  (without the consideration of double clicks)
\begin{eqnarray}
&&H_{min}(Z|E)~~~~~~~~\nonumber\\&&=-\log_{2}(\max_i P_{i}),\nonumber\\&&=-\log_{2}\{\max[\frac{p_{0}^{(1)}(1-p_{1}^{(0)})}{\textit{Q}_{01}},\frac{p_{1}^{(0)}(1-p_{0}^{(1)})}{\textit{Q}_{01}}]\},\nonumber\\
&&\textit{Q}_{01}=p_{0}^{(1)}(1-p_{1}^{(0)})+p_{1}^{(0)}(1-p_{0}^{(1)}).
\end{eqnarray}
Without loss of generality, we consider the parameter difference of the two detectors, and $ H_{min}(Z|E) $ will become:

	\begin{eqnarray}
&&	H_{min}(Z|E)\nonumber\\&&=-\log_{2}\{\max_{\{\alpha,\beta\},\{m,n\}=\{1,0\}}[\frac{p_{\alpha}^{(m)}(1-p_{\beta}^{(n)})}{\textit{Q}_{\alpha\beta}^{mn}}]\},\nonumber\\
	&&\textit{Q}_{\alpha\beta}^{mn}=p_{\alpha}^{(m)}(1-p_{\beta}^{(n)})+p_{\beta}^{(n)}(1-p_{\alpha}^{(m)}),\nonumber\\
	&&p_{\alpha}^{(1)}=1-\tau_{\alpha}(1-e_{d\alpha})(1-\frac{\hat{p_{\alpha}}}{1-\hat{p_{\alpha}}}),\nonumber\\
	&&p_{\alpha}^{(0)}=1-\tau_{\alpha}(1-e_{d\alpha}).\label{10a}
	\end{eqnarray}

In what follows, we discuss the total conditional minimum entropy $ H_{min}(A|E) $ in SI-QRNGs. In a typical SI-QRNG, a key idea is that the protocol with randomness generation and randomness extraction can be seen as similar to that with error correction and randomness generation\cite{cao2016source}, which borrows a similar technique from the security analysis of the QKD\cite{shor2000simple}. Corresponding to the original protocol discussed in the previous section, the equivalent virtual protocol can be described as follows: the input states $ \rho $ will be corrected to perfect diagonal states $ \ket{+} $ by a phase error correction with losses of the $ h(e_{bx}) $ states and the remaining $ 1-h(e_{bx}) $ corrected states $ \ket{+} $ can be used to generate perfect random numbers in the Z basis. In this sense, when we estimate $ H_{min}(A|E) $, it is reasonable that we first estimate the influence of the error rate $ e_{bx} $ in the X basis and correct all the states to $ \ket{+} $, and then the problem can be changed to the estimation of the conditional minimum entropy $ H_{min}(Z|E) $ with these input states $ \ket{+} $. In the previous section, we obtained the error rate $ EQ $ in the X basis with our afterpulse model in Eq.~(\ref{4c}). Therefore, the number of corrected states $ \ket{+} $ is:
\begin{eqnarray}
n_{corrected}=n_{z}(1-h(EQ)),
\end{eqnarray}
where $ n_{z} $ is the number of pulses detected in the Z basis. 

Note that the double clicks in the Z basis should also be considered. A series of true random numbers will be input to fill in these double-click events, and, of course, these true random numbers will be subtracted from the final bit rate. According to the probabilities of single click and double-click events $ Q_{single} $, and $ Q_{double} $ in the Z basis in Eqs.~(\ref{4a}), and (\ref{4b}) and the conditional minimum entropy in the Z basis $ H_{min}(Z|E) $ in Eq.~(\ref{10a}), we can obtain $ H_{min}(A|E) $ by:
\begin{eqnarray}
H_{min}(A|E)&=&[H_{min}(Z|E)Q_{single}+1\times Q_{double}]\nonumber\\&&\times[1-h(EQ)]-{Q_{double}}.\label{HAE} \label{12}
\end{eqnarray}
 
In the previous discussion, we considered the afterpulse probability $ P_{ap\alpha} $ with Eq.~(\ref{3}) under the condition that there are infinite former responses\cite{fan2018afterpulse}. However, in real implementations, there are finite pulses in front and hence only these corresponding responses' afterpulse contributions should be calculated. Here we consider this condition and recalculate the afterpulse probability $ P^{'}_{ap\alpha} $ with finite former responses.

Here, we assume that there are $ m $ previous detection windows. As in our earlier definition, $ \hat{p}_{j\alpha} $ is the first-order afterpulse probability coefficient contributed by the former \textit{j}th detection. A high-order afterpulse is the superposition of the contributions of each first-order afterpulse. All of these high-order afterpulse probabilities contributed by the former \textit{k}th detection window $ p^{k}_{ap\alpha} $ are given by:
\begin{eqnarray}
p^{k}_{ap\alpha}=\sum_{\sum i=k}\prod_{i}\hat{p}_{i\alpha}.
\end{eqnarray}
According to the previous analysis, always letting one of the detectors respond before is the best case for Eve. The total afterpulse probability $ P^{'}_{ap\alpha}(m) $ in this case changes to
\begin{eqnarray}
P^{'}_{ap\alpha}(m)=\sum_{k=1}^{m} p^{k}_{ap\alpha}=\sum_{k=1}^{m} \sum_{\sum i=k}\prod_{i}\hat{p}_{i\alpha}.\label{14}
\end{eqnarray}

Previous works\cite{cova1991trapping}\cite{korzh2015afterpulsing} show that the afterpulse probability $ \hat{p}_{j\alpha} $ at time $ t $ conforms to an \textquotedblleft exponential model\textquotedblright where the characteristic decay of the afterpulse probability depends on the depth of the levels in which the charges are trapped. A simplified model for the gating detector can be given by:
\begin{eqnarray}
\hat{p}_{j\alpha}=A_{\alpha}e^{-m\omega_{\alpha}},\label{15}
\end{eqnarray}
where $ \omega_{\alpha} $ is the ratio between the gating time and the de-trapping lifetime, and $ A_{\alpha} $ is the amplitude factor for the depth level. Therefore, the total afterpulse probability contributed by $ m $ previous detection windows can be derived as:
\begin{eqnarray}
P^{'}_{ap\alpha}(m)=A_{\alpha}e^{-\omega_{\alpha}}\frac{[(1+A_{\alpha})e^{-\omega_{\alpha}}]^{m}-1}{(1+A_{\alpha})e^{-\omega_{\alpha}}-1},\label{16}
\end{eqnarray}
and $ \omega_{\alpha}>\ln(1+A_{\alpha}) $ is the convergence condition for $ m \rightarrow \infty $. In the METHODS section, we will show the details of the derivation and the relation to infinite previous pulses.

\begin{figure}[htbp]
	\centering
	\includegraphics[width=0.9\linewidth]{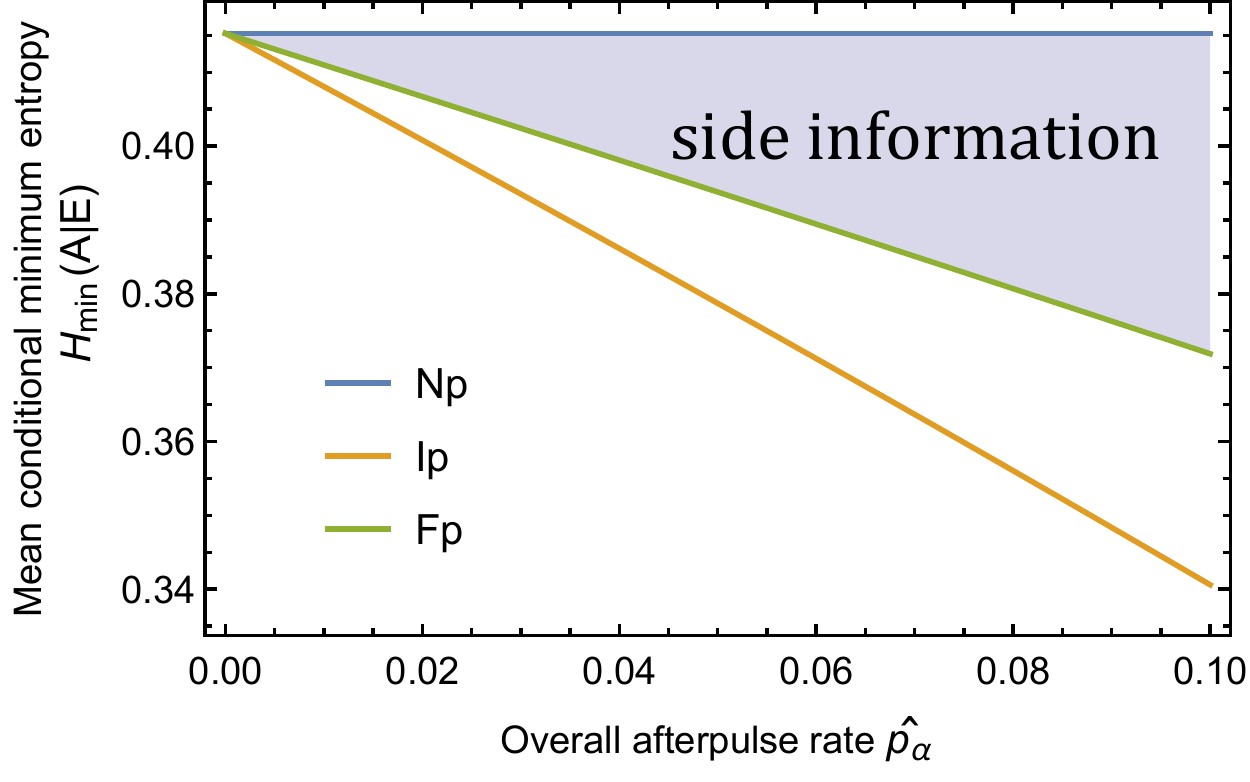}
	\caption{The relation between $ H_{min}(A|E) $ and $ \hat{p_{\alpha}} $ with no afterpulse (Np), previous infinite afterpulse (Ip) and previous finite afterpulse (Fp), which consists of 1000 pulses. The factors $ \omega_{\alpha}=0.001 $ and $ A_{\alpha} $ are related to $ \hat{p_{\alpha}} $. The pulses with coherent states that contain $ \nu=10 $ photons insert SPDs with $ \eta_{\alpha}=0.1 $ and $ e_{d}=6\times10^{-7} $. The shadow gap is the side information leaked to Eve due to the afterpulse.}
	\label{fig:p-r}
\end{figure}

In Fig.\ref{fig:p-r}, we have shown the relation between the conditional minimum entropy $ H_{min}(A|E) $ and the overall afterpulse rate $ \hat{p_{\alpha}} $ with and without the afterpulse. The two lines of $ H_{min}(A|E) $ with the afterpulse have a significant inverse relationship with $ \hat{p_{\alpha}} $ and decrease nearly $ 20\% $ when $ \hat{p_{\alpha}} $ increases to 0.1. We can also see that the conditional minimum entropy with previous infinite pulses will decrease more rapidly than that with previous finite pulses. This means that limiting the number of responses in a detection period is an effective way to enhance the conditional minimum entropy, especially under the condition with the high afterpulse probability.

\hfill

\noindent{\textit{Detection efficiency and the photon distribution}}

\begin{figure}[htbp]
	\centering
	\includegraphics[width=0.9\linewidth]{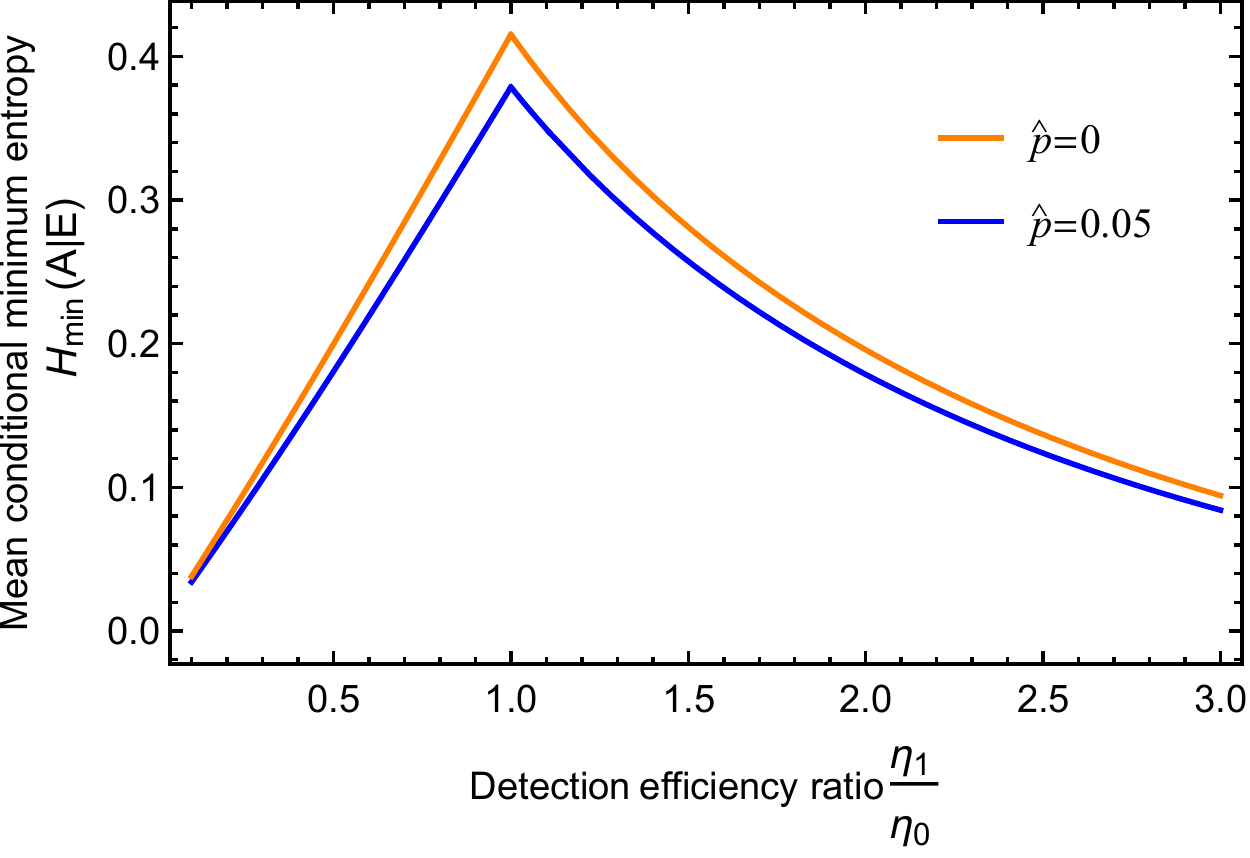}
	\caption{The relation between $ H_{min}(A|E) $ and $ \frac{\eta_{1}}{\eta_{0}} $ with different $ \hat{p_{\alpha}} $ (0, 0.05). The pulses with a Poisson distribution that contains $ \nu=10 $ photons insert SPDs with $ \eta_{\alpha}=0.1 $ and $ e_{d}=6\times10^{-7} $. }
	\label{fig:yita-r}
\end{figure}

In addition to the afterpulse, the other non-negligible parameter that might affect the conditional minimum entropy is the zero photon distribution probability $\tau_{\alpha} $ in Eq.~(\ref{2}). In the SI-QRNG protocol, $\tau_{\alpha} $ is determined by the detection efficiency, loss, and photon distribution input to the detector. The difference in detection efficiency and loss will cause an imbalance in the final random numbers and have a severe impact on the security of the scheme. In Fig.\ref{fig:yita-r}, we show that with detection efficiency mismatch, $ H_{min}(A|E) $ falls sharply. Moreover, the photon distribution input to the detector also influences the conditional minimum entropy. In previous works, the photon distribution is usually seen as a Poisson distribution, which is not universal for sources and might result in information leakage as the afterpulse. Here, we analyse how the detection efficiency, loss, and photon distribution affect $\tau_{\alpha} $ and design a scheme to monitor these influences.

In theory, the photon distributions before and after loss satisfy a Bernoulli transformation. The untrusted photon source through a filter, which is used to guarantee the source in single mode, will become a photon number mixed state by a phase randomizer\cite{lo2005decoy}:
\begin{eqnarray}
\rho=\sum_{n=0}^{\infty}P_{untrusted}(n)\ket{n}\bra{n},
\end{eqnarray}
and then through the loss transmittance in the system, $ t_{all\alpha} $ and the detector's efficiency, $ \eta_{\alpha} $, the photon distributions input to the detectors will become\cite{lee1993external}
\begin{eqnarray}
D(m)&=&B[P_{untrusted}(n),\xi_{\alpha}]\nonumber\\&=&\sum_{n=m}^{\infty}P_{untrusted}(n)\binom{n}{m}\xi_{\alpha}^{m}(1-\xi_{\alpha})^{n-m},
\end{eqnarray}
where $ \xi_{\alpha}= t_{all\alpha}\eta_{\alpha} $. As a consequence, we can obtain the zero photon distribution probability $\tau_{\alpha} $ by
\begin{eqnarray}
\tau_{\alpha}&=&D(0)\nonumber\\&=&\sum_{n=0}^{\infty}P_{untrusted}(n)(1-\xi_{\alpha})^{n}.\label{tau}
\end{eqnarray}

Now, the key question for Alice becomes how to monitor the distribution of the untrusted source $ P_{untrusted}(n) $. With the existence of the afterpulse, it is difficult to obtain the distribution precisely only through the SPDs. Fortunately, this can be done by borrowing a similar technique from the source monitor of the QKD\cite{xu2010passive}. As shown in Fig.\ref{fig:fig1}, in the photon distribution monitor block, a beam splitter (BS) is used to take out a beam of photons to a photodetector (PD) which is used to monitor the photon distribution of the source. Then the others will go through the BS, and an attenuation with the attenuation coefficient $ t_{0} $ is placed to guarantee that the distribution after it is the same as that detected in the PD, which satisfies:  
\begin{eqnarray}
t_{0}=\frac{(1-\eta_{BS})}{\eta_{BS}}\eta_{DET},
\end{eqnarray}
where $ \eta_{BS} $ is the transmittance of the BS and $ \eta_{DET} $ is the detection efficiency of the PD. Through random sampling of the pulses, we can estimate the photon distribution $ P_{untrusted}(n) $, and with Eq.~(\ref{tau}), the zero photon distribution probability $\tau_{\alpha} $ after the attenuation rate $ t_{all\alpha} $ and the detection efficiency $ \eta_{\alpha} $ can be obtained, which can help us to precisely estimate $ H_{min}(A|E) $ with Eq.~(\ref{HAE}). And a series of biased random seeds will be consumed in the random sampling of the pulses, which guarantees that the distribution measurements are independent.

\hfill

\noindent{\textit{Simulation in finite-size regime}}

\noindent In practice, the resources of Alice are limited and the system can run for only finite time. Limited samplings will suffer from statistical fluctuations, which might enable attacks by Eve. Therefore, it is of great importance to estimate the parameters in the finite-size regime for the final random number security. Here, we consider the influence of the finite data size on the error estimation in the X basis as well as the process of photon distribution monitoring. We also consider the composable security and obtain the final random number rate $ R_{f} $ with the total security parameter $ \zeta $.

In the error parameter estimation step, Alice can obtain $ EQ $ in the X basis according to Eq.~(\ref{4c}) and can approximate the phase error rate $ e_{pz} $ in the Z basis by $ EQ $. However, due to statistical fluctuations,  $ e_{pz} $ cannot be estimated accurately and the method of approximating it is crucial. In this section, we use two approaches to bound it: one of the methods is random sampling theory used in \cite{cao2016source} and the other is entropy inequality.

\begin{figure*}[htbp]
	\centering
	\includegraphics[width=0.9\textwidth]{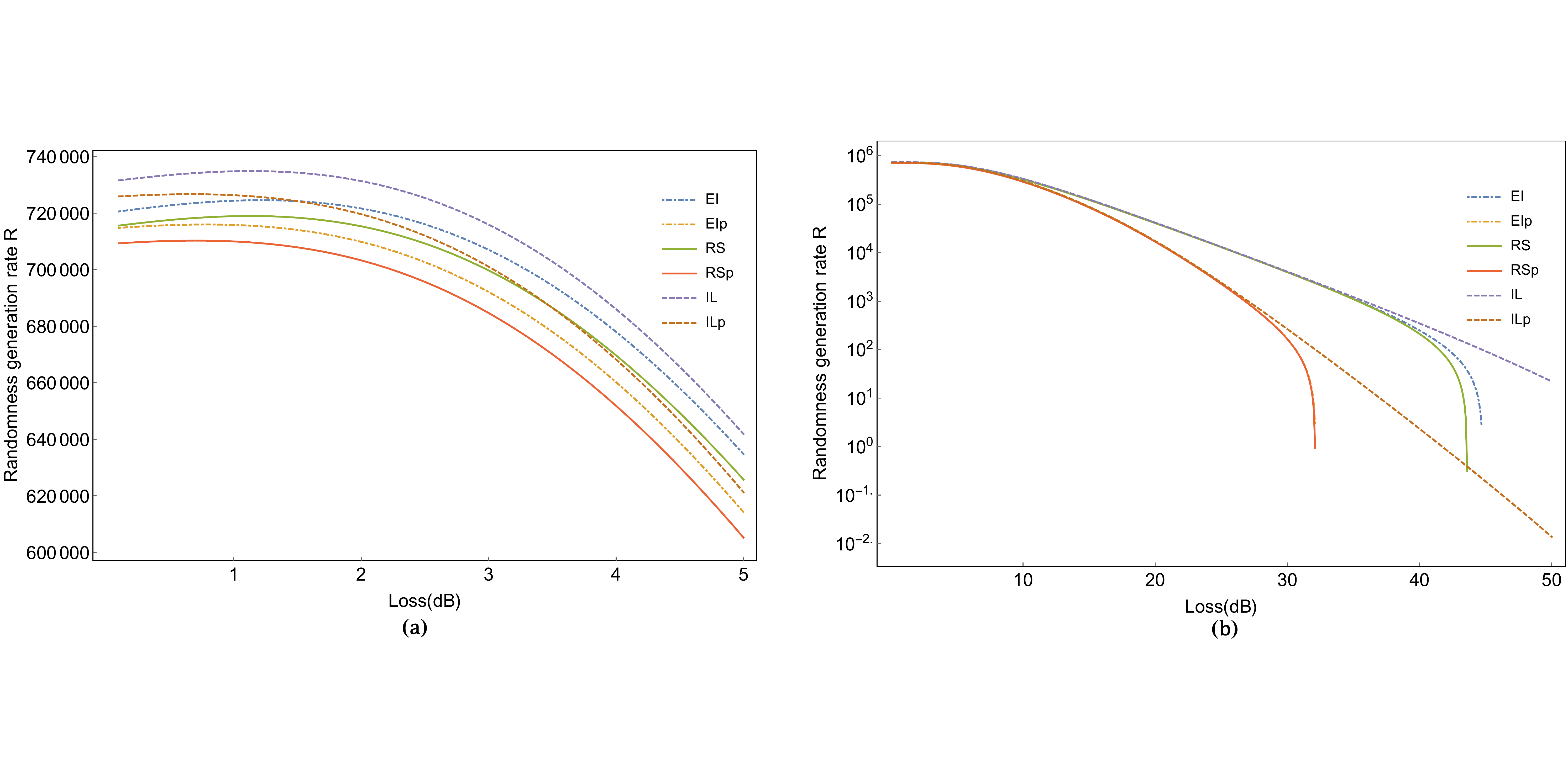}\label{fig:mu-ra}
	\caption{Optimal randomness generation rates as a function of the loss of VOA with different values of $ \hat{p}_{\alpha} $ and error estimation methods including random sampling (RS), entropy inequality (EI) and infinite length (IL). RS, EI and IL represent the models without the afterpulse and RSP, EIP and ILP represent the models with the afterpulse $\hat{p}_{\alpha}=5\% $. The experimental parameters are listed in Table.\ref{tab:my-table}, and we assume that the pulse distribution is the coherent state that contains $ \nu=50 $ photons initially. Fig.(a) shows the condition with low loss and Fig.(b) shows that with the loss of up to 50 dB.}
	\label{fig:mu-r}
\end{figure*}

The upper bound of $ e_{pz} $ can be defined by
\begin{eqnarray}
e_{pz}\leqslant EQ+\theta,~~~~~~
\end{eqnarray}
and on the basis of the random sampling analysis in \cite{fung2010practical}, $ \theta $ is the deviation due to statistical fluctuations which is bounded by
\begin{eqnarray}
~~~~~\varepsilon_{e}&&=Prob(e_{pz}>EQ+\theta)\nonumber\\&&\leqslant \frac{1}{\sqrt{q_{x}(1-q_{x})EQ(1-EQ)N}}2^{-n\zeta(\theta)},\label{theta1}
\end{eqnarray}
where $ \zeta(\theta)=h(EQ+\theta-q_{x}\theta)-q_{x}h(EQ)-(1-q_{x})h(EQ+\theta) $, $ q_{x} $ is the rate of X basis measurement and $ N $ is the total number of pulses. With the model presented in the previous sections, the number of final random bits is given by
\begin{eqnarray}
R&=&n_{z}[(H_{min}(Z|E)Q_{single}+1\times Q_{double})\nonumber\\&&\times(1-h(EQ+\theta))-{Q_{double}}] -t_{e},\label{r}
\end{eqnarray}
where $ n_{z} $ is the number of pulses measurement in Z basis and $ \varepsilon=2^{-t_{e}} $ is the failure probability of the randomness extraction which satisfies the relation with security parameter $ \varepsilon_{all}=\sqrt{(\varepsilon_{e}+2^{-t_{e}})(2-\varepsilon_{e}-2^{-t_{e}})}$.

Furthermore, the entropy inequality is an alternative method to bound the final random bit rate through bounding  $ e_{pz} $ \cite{tomamichel2012tight} by:
\begin{eqnarray}
\theta=\sqrt{\frac{n_{z}+n_{x}}{n_{z}n_{x}}\frac{n_{x}+1}{n_{x}}\ln\frac{2}{\varepsilon_{e}}}.\label{theta}
\end{eqnarray}
Here we set $ \varepsilon_{e} $ as the total security parameter $ \varepsilon_{all} $ because there is no error correction, which is different from the QKD. The final random bit rate can be bounded by
\begin{eqnarray}
R&=&n_{z}[(H_{min}(Z|E)Q_{single}+1\times Q_{double})\nonumber\\&&\times(1-h(EQ+\theta))-{Q_{double}}] -2\log_{2}\frac{1}{\varepsilon_{all}}.\label{r2}
\end{eqnarray}

In what follows, we present and discuss the results of the numerical simulation. We use the experimental parameters listed in Table.\ref{tab:my-table}. The relations between the loss of VOA and the randomness generation rates, with different values of $ \hat{p}_{\alpha} $ and different methods, are shown in Fig.\ref{fig:mu-r}. Compared with the rate without the afterpulse, the rate with the afterpulse is lower and decreases more obviously; from the loss not exceeding 5 dB, the influence of the afterpulse has become more memorable than that of statistical fluctuation gradually and the afterpulse will result in the lower tolerance for fewer photons as well as more information leakage to Eve. Moreover, the bound of the entropy inequality results in a higher randomness generation rate than random sampling. The rate peak is approximately 1\textasciitilde2 dB, and there is a slight difference with different analysis methods.

	\begin{table}[]
	\centering
	\caption{List of the experiment parameters used in numerical simulations. $ N $ is the total number of pulses. $ e_{q} $ is the misalignment-error probability. $ v $ is the rate of pulses measured in Z basis.}
	\renewcommand{\arraystretch}{1.5}
	\begin{ruledtabular}
		\begin{tabular}{cccccccccl}
			\textrm{$N$}       & \textrm{ $ \varepsilon_{all} $} & \textrm{ $ \varepsilon_{d}$} & \textrm{$ e_{d} $} & \textrm{$ \eta_{\alpha} $ } & \textrm{$ \eta_{BS} $} & \textrm{$ e_{q} $} & \textrm{$ v $} & \textrm{$ t_{e} $} & \textrm{$ q_{x} $} \\
			\colrule
			$ 10^{10} $ &       $  2\times2^{-50} $      &       $  2^{-50} $  & $ 6\times10^{-7} $ &            10\%             &            0.5             &        0.02         &   $ 10^{6}$    &        100         & 0.02
		\end{tabular}
	\end{ruledtabular}
	\label{tab:my-table}
\end{table}

\begin{figure}[htbp]
	\centering
	\includegraphics[width=0.9\linewidth]{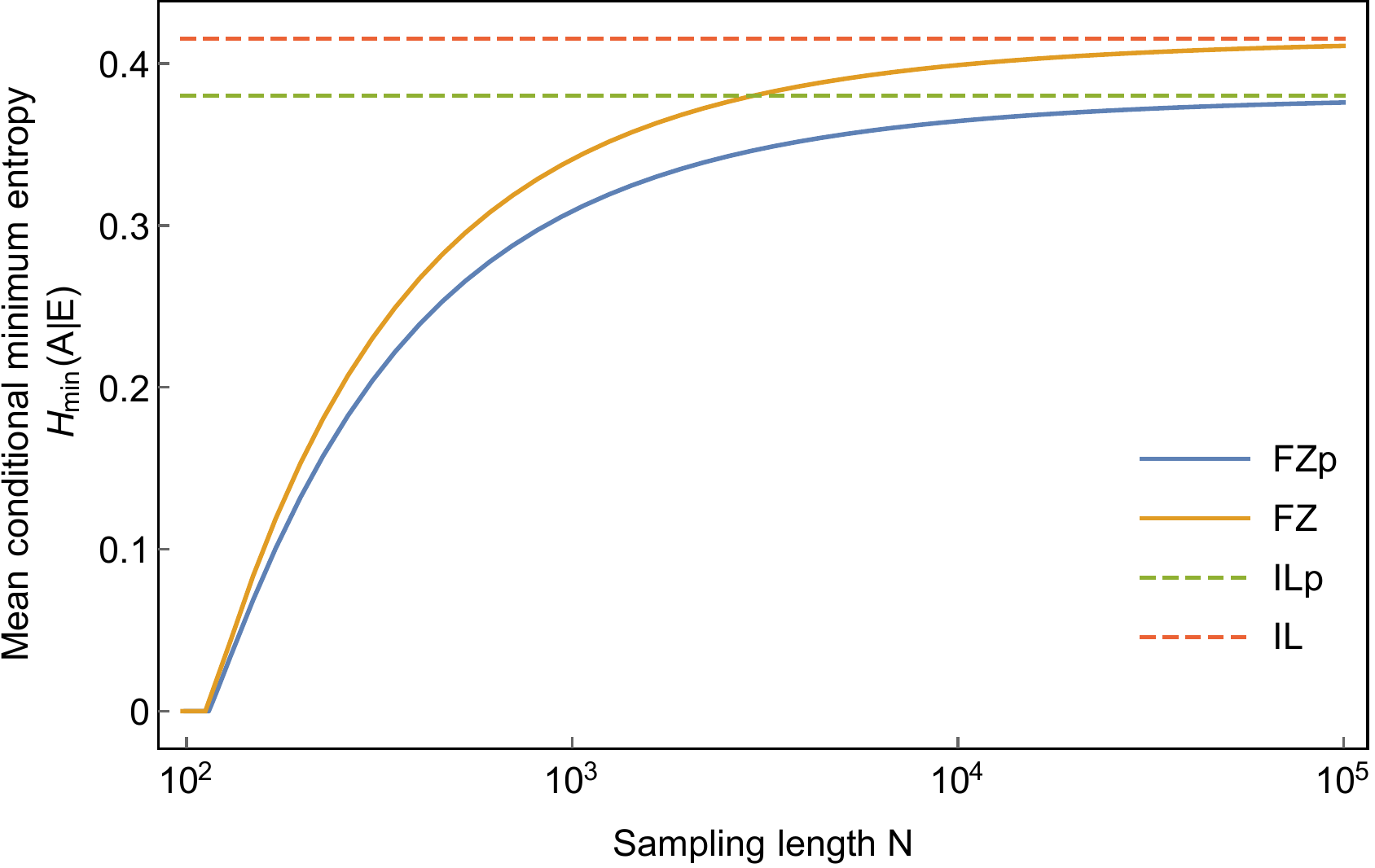}
	\caption{Conditional minimum entropy $ H_{min}(A|E) $ as a function of the sampling length with the finite size (FS) and infinite length (IL).  FZ and IL represent the rates without the afterpulse, and FZP and ILP represent the rates with the afterpulse $\hat{p}_{\alpha}=5\% $. Assume that pulses with coherent states contain $ \nu=10 $ photons.}
	\label{fig:d0-r}
\end{figure}

Furthermore, there is also statistical fluctuation in the photon distribution monitor. Even if we use an ideal photodetector, the distribution estimation will also fluctuate with limited sampling pulses. Assume that Alice randomly chooses $ N $ pulses in the photon distribution monitor, and the vacuum probability value input to SPDs $ \tau'_{\alpha} $ can be estimated by Eq.~(\ref{tau}). According to Hoeffding’s inequality\cite{hoeffding1963probability}, the confidence interval of the vacuum probability is $ \tau_{\alpha}\in[\tau'_{\alpha}-\delta_{d},\tau'_{\alpha}+\delta_{d}] $ with confidence level $  \kappa=1-\varepsilon_{d} $, where $ \varepsilon_{d} $ is the distribution estimation failing probability, which is given by $ \varepsilon_{d}=2exp(-2N\delta_{d}^{2}) $. To show the effect of statistical fluctuations in the numerical simulation, we assume that the untrusted source distribution estimation result is a Poissonian distribution. As shown in Fig.\ref{fig:d0-r}, the limited random sampling pulses will cause the large gap between the ideal $ H_{min}(A|E) $ and practical $ H_{min}(A|E) $ with finite size effect and it will be shrunken when the length of the random sampling pulses is at least $ 10^{5} $.

For a larger system, giving one monolithic security proof is error prone. Therefore, in the past few years, composable security, as a solution to this problem, has been developed in the research of QRNGs\cite{drahi2019certified,avesani2018source} as well as QKDs\cite{fung2010practical,tomamichel2012tight,muller2009composability,lim2014concise}. Here we also establish composable security and obtain the final random number rate $ R_{f} $.

In general, the raw random sequence of Alice can be quantum correlated with a quantum state that is held by Eve. Mathematically, this situation is described by the classical quantum state
\begin{eqnarray}
\rho_{AE}=\sum_{i}p_{i}\ket{i}\bra{i}\otimes\rho^{i}_{E},
\end{eqnarray}
where $ \{\ket{i}\} $ denotes an orthonormal basis for Alice’s system, and the subscript E indicates the system of Eve. It is easy to see that, for any attack, the state resulting from the run of a perfectly secure scheme has the form $ \rho'_{AE}=\rho_{A}\otimes\rho_{E} $ where $ \rho_{A}=\sum_{i}\frac{1}{\mid I\mid}\ket{i}\bra{i} $ is the uniform mixture of all possible values of the bit string. As is common in quantum cryptography, a QRNG protocol is $ \zeta $-secret if and only if, for any attack, the classical quantum state $ \rho_{AE} $ satisfies
\begin{eqnarray}
\frac{1}{2}\parallel\rho_{AE}-\rho'_{AE}\parallel_{1}\leqslant\zeta,
\end{eqnarray}
where $ \parallel\cdot\parallel_{1}$ denotes the trace norm. In our protocol, the total failing probability is a combination of two processes, error estimation and photon distribution estimation. As the composable security analysis in the QKD\cite{fung2010practical}, $ \zeta $ can be given by
\begin{eqnarray}
\zeta\leqslant\sqrt{(\varepsilon_{d}+\varepsilon_{e}+2^{-t_{e}})(2-\varepsilon_{d}-\varepsilon_{e}-2^{-t_{e}})}, \label{var}
\end{eqnarray}
where $ \varepsilon_{e} $ and $ \varepsilon_{d} $ are the failure probabilities of the error estimation and photon distribution estimation, respectively. $ 2^{-t_{e}} $ is the failure probability of the randomness extraction. Combining the analysis in Eq.~(\ref{var}), the final random number rate is
\begin{eqnarray}
R_{f}&=&\min_{\tau_{\alpha}\in[\tau'_{\alpha}-\delta_{d},\tau'_{\alpha}+\delta_{d}]} \{n_{z}[(H_{min}(Z|E)Q_{single}+\nonumber\\&&1\times Q_{double})\times(1-h(EQ+\theta))-{Q_{double}}]\}\nonumber\\&& -t_{e},
\end{eqnarray}
where $ \theta $ is the deviation due to statistical fluctuations, which is bounded by Eq.~(\ref{theta1}).

\hfill

\noindent{\bf DISCUSSION}

\noindent In conclusion, we have pointed out and evaluated the security loopholes of practical imperfect devices in discrete-variable SI-QRNGs. By our analysis, Eve might obtain the imperfection parameters of the measurement and obtain side information to enhance the guessing probability of outcomes. The entropy estimation error can reach 20$\%$ without consideration of the large afterpulse and even higher under the conditions of the large mismatch of the detector efficiency and the misestimation of the photon number distribution. To solve these problems, we provide a new protocol with distribution monitoring and entropy estimation methods to extract the secure randomness bits with these existing factors. By analysing the finite-size effect, we show the final randomness rates and find that under some conditions, the influence of the afterpulse will exceed that of the finite size effect. Finally, we establish a composable security model to guarantee the security of the total protocol. Compared with the general SI-QRNG in \cite{cao2016source}, our model is more practical and more compatible with imperfect devices.

Our model also provides a way to achieve high-speed SI-QRNGs. The common discrete variable SI-QRNGs are usually limited by the frequency of SPDs which is approximately 100 Mbps. A faster counting rate is not allowed due to the leaked information of the afterpulse. With our model, however, the afterpulse is also a factor in the randomness rate, which means that we can use SPDs with higher afterpulse rates. In recent research, a SPD, with a rate of up to 500 MHz has been presented, which has an afterpulse rate of nearly 10\%\cite{comandar2015gigahertz}. This is not tolerated in most protocols but with our analysis, a high-speed SI-QRNG can be obtained without any security problem of the afterpulse. Therefore, with the promotion rates of SPDs, the imperfect-device-compatible model will be more suitable for high-speed scenarios, and this may make it potentially easier for a discrete-variable GHz-SI-QRNG to be realized.

\hfill

\noindent{\bf METHODS}

\noindent{\bf Calculation of the autocorrelation coefficient}

\noindent In this section, we will show the details of calculating $ a_{i}^{p} $. According to Eq.~(\ref{5}), we have given out the general expression of the autocorrelation coefficient. To obtain the final result, it is important to derive each statistical value of the numerator and denominator.

Here we assume that the n-bits raw random number sequence contains $ nk $ bits with the value '0' and $ n(1-k) $ bits with the value '1'. The expectation and variance can be given by
\begin{eqnarray}
\bar{x}&&=k,\nonumber\\
\sum_{j=1}^{n}(x_{j}-\bar{x})^{2}&&=nk(1-k)^{2}+n(1-k)k^{2}.
\end{eqnarray}

Now we consider the value of the numerator $ \sum_{j=1}^{n-i}(x_{j}-\bar{x})(x_{j+i}-\bar{x}) $. For the \textit{j}th resultful detection event, considering that $ x_{j}=1 $ or $ 0 $, the afterpulse probability of detector $ D_{\alpha} $ ($ \alpha\in\{0,1\} $) in the (\textit{j}+\textit{i})th resultful detection event will become (where the high-order afterpulse generated by the \textit{j}th response is disregarded):
\begin{eqnarray}
&& P_{ap\alpha}^{i1}=\frac{\hat{p_{\alpha}}}{1-\hat{p_{\alpha}}}p^{d}_{b\alpha}+\hat{p}_{i\alpha}(1-p^{d}_{b\alpha}),\nonumber\\
or~~&& P_{ap\alpha}^{i0}=\frac{\hat{p_{\alpha}}}{1-\hat{p_{\alpha}}}p^{d}_{b\alpha}+\hat{p}_{i\alpha}(-p^{d}_{b\alpha}),\label{28}
\end{eqnarray}
and the corresponding response probability is
\begin{eqnarray}
&&p_{\alpha}^{i1}=1-\tau_{\alpha}(1-e_{d\alpha})(1-P_{ap\alpha}^{i1}),\nonumber
\\
or~~&&
p_{\alpha}^{i0}=1-\tau_{\alpha}(1-e_{d\alpha})(1-P_{ap\alpha}^{i0}).\label{29}
\end{eqnarray}
Therefore, the \textit{prior} statistical autocorrelation coefficient can be derived by ($ n\gg i $):
\begin{eqnarray}
a^{p}_{i}=&&\frac{\sum_{j=1}^{n-i}(x_{j}-\bar{x})(x_{j+i}-\bar{x})}{\sum_{j=1}^{n}(x_{j}-\bar{x})^{2}}\nonumber\\
=&&\frac{\sum_{x_{j}=1}^{j}(1-\bar{x})(x_{j+i}-\bar{x})+\sum_{x_{j}=0}^{j}(0-\bar{x})(x_{j+i}-\bar{x})}{\sum_{j=1}^{n}(x_{j}-\bar{x})^{2}}\nonumber\\
=&&\dfrac{1}{nk(1-k)^{2}+n(1-k)k^{2}}\nonumber\\
&&\times\{(n-i)k(1-k)\nonumber\\
&&\times[(1-k)p^{i1}_{1}(1-p^{i0}_{0})+(0-k)p^{i0}_{0}(1-p^{i1}_{1})]\nonumber\\
&&+(n-i)(1-k)(-k)\nonumber\\
&&\times[(1-k)p^{i0}_{1}(1-p^{i1}_{0})+(0-k)p^{i1}_{0}(1-p^{i0}_{1})]\}\nonumber\\
=&&[p^{i1}_{1}(1-p^{i0}_{0})-p^{i0}_{1}(1-p^{i1}_{0})](1-k)\nonumber\\
&&+[p^{i0}_{0}(1-p^{i1}_{1})-p^{i1}_{0}(1-p^{i0}_{1})](-k).
\end{eqnarray}

According to Eqs.~(\ref{28}) and (\ref{29}), the response probabilities $ p_{\alpha}^{i1} $ and $ p_{\alpha}^{i0} $ are both linear about $ \hat{p}_{i\alpha} $. Therefore, it is obvious that $ a^{p}_{i} $ is a quadratic function about $ \hat{p}_{i\alpha} $ and when we choose the same parameters for the two detectors, $ a^{p}_{i} $ will become:
\begin{eqnarray}
a^{p}_{i}=&&[p^{i1}_{0}(1-p^{i0}_{0})-p^{i0}_{0}(1-p^{i1}_{0})](1-k)\nonumber\\
&&+[p^{i0}_{0}(1-p^{i1}_{0})-p^{i1}_{0}(1-p^{i0}_{0})](-k)\nonumber\\
=&&p^{i1}_{0}-p^{i0}_{0}\nonumber\\
=&&\tau_{0}(1-e_{d0})p^{d}_{b0}\hat{p}_{i0}.
\end{eqnarray}
and it will degenerate to a linear function about $ \hat{p}_{i\alpha} $.

\hfill

\noindent{\bf Calculation of the total afterpulse probability for previous finite responses}

\noindent In this section, we will derive the total afterpulse probability. According to Eqs.~(\ref{14}) and (\ref{15}), we can obtain the $ P^{'}_{ap\alpha}(m) $ by:
\begin{eqnarray}
P^{'}_{ap\alpha}(m)&&=\sum_{k=1}^{m} \sum_{\sum i=k}\prod_{i}\hat{p}_{i\alpha}\nonumber\\
&&=\sum_{k=1}^{m} \sum_{\sum i=k}\prod_{i}A_{\alpha}e^{-i\omega_{\alpha}}\nonumber\\
&&=\sum_{k=1}^{m} \sum_{j=1}^{k}\binom{k}{j}A_{\alpha}^{j}e^{-k\omega_{\alpha}}\nonumber\\
&&=\sum_{k=1}^{m} A_{\alpha}(1+A_{\alpha})^{k-1}e^{-k\omega_{\alpha}}\nonumber\\
&&=A_{\alpha}e^{-\omega_{\alpha}}\frac{[(1+A_{\alpha})e^{-\omega_{\alpha}}]^{m}-1}{(1+A_{\alpha})e^{-\omega_{\alpha}}-1}.\label{b1}
\end{eqnarray}
To compare the afterpulse probability for previous infinite responses $ P_{ap\alpha}(all) $, it is necessary to obtain $ \hat{p_{\alpha}} $ by:
\begin{eqnarray}
\hat{p_{\alpha}}&&= \sum^{m}_{j=1}\hat{p}_{j\alpha}= \sum^{m}_{j=1}A_{\alpha}e^{-j\omega_{\alpha}}
=A_{\alpha}e^{\omega_{\alpha}}\frac{e^{-m\omega_{\alpha}}-1}{e^{-\omega_{\alpha}}-1}\nonumber\\
&&=\frac{A_{\alpha}e^{-\omega_{\alpha}}}{1-e^{-\omega_{\alpha}}}~~~(m \rightarrow \infty),
\end{eqnarray}
and then according to Eq.~(\ref{3}), we can get $ P_{ap\alpha}(all) $ by:
\begin{eqnarray}
P_{ap\alpha}(all)= \frac{\hat{p_{\alpha}}}{1-\hat{p_{\alpha}}}=\frac{A_{\alpha}e^{-\omega_{\alpha}}}{1-(1+A_{\alpha})e^{-\omega_{\alpha}}}.
\end{eqnarray}

It is obvious that $ P^{'}_{ap\alpha}(m) $ is a monotonic decreasing function and will degenerate to $ P_{ap\alpha}(all) $ when $ m \rightarrow \infty $, and the condition of convergence for Eq.~(\ref{b1}) is $ (1+A_{\alpha})e^{-\omega_{\alpha}}<1 $, that $ \omega_{\alpha}>\ln(1+A_{\alpha}) $.

\hfill

\noindent {\bf Data availability}

\noindent The data that support the findings of this study are available from the corresponding author upon reasonable
request.

\hfill

\noindent{\bf ACKNOWLEDGEMENTS}

\noindent We thank Xiongfeng Ma and Xingjian Zhang for helpful discussions. This work was supported by the National
Key Research And Development Program of China
(Grant No. 2018YFA0306400), the National Natural Science
Foundation of China (Grants No. 61622506, No.
61575183, 61627820, No. 61475148, and No. 61675189),  and the Anhui Initiative in Quantum
Information Technologies.

\hfill

\noindent{\bf Competing interests}

\noindent The authors declare that they have no competing interests.

\hfill

\noindent{\bf AUTHOR CONTRIBUTIONS}

\noindent S.W., Z-Q.Y., G-J.F-Y., R.W. and X.L. conceived the basic idea of the protocol. X.L. finished the details of the protocol and the simulations. S.W. and X.L. wrote the paper. S.W., W.C., D-Y.H., Z.Z., G-C.G. and Z-F.H. supervised the project and all authors participated in discussions.

\hfill

\noindent{\bf REFERENCES}
\nocite{*}

\bibliography{apssamp}% Produces the bibliography via BibTeX.

\end{document}